\documentstyle[11pt,twoside,psfig]{article}
\begin{document}
\title{{\bf Comment on a relativistic model for coalescing neutron 
star binaries}}
\author{R. Rieth and G. Sch\"afer \\
\\
Max--Planck--Gesellschaft,
Arbeitsgruppe Gravitationstheorie an der Universit\"at Jena,\\
Max--Wien--Platz 1, D--07743 Jena, Germany}
\maketitle
\begin{abstract}
The Wilson approximate dynamics and the Einstein dynamics
are compared for binary systems. At the second post-Newtonian approximation, genuine 
two-body aspects are found to differ by up to 114\%. In the regime of a formal innermost
stable circular orbit (ISCO) the both dynamics differ by up to 7\%.
\end{abstract}
Recently Wilson and Mathews [1] proposed a truncated version of
the Einstein field equations to treat the coalescence of binary neutron stars
in a much simplified but still sufficiently precise manner. The main idea of
this approximation is to
neglect the independent (``true'') degrees of freedom of the gravitational field,
i.e. in particular, the full gravitational radiation content. For spherically
symmetric processes the proposed scheme is identical with the Einstein
equations, in non-spherically symmetric dynamical situations,
even stationary ones, the proposed scheme and the Einstein theory only coincide
at the first post-Newtonian level of approximation. It is perhaps worth
mentioning that, in contrast to the Einstein theory, the proposed scheme should
allow post-Newtonian series expansions to all orders in (integer) powers of
$1/c^2$.

Most recently, Wilson et al. [2] applied the Wilson
scheme to the question of instabilities in close neutron star binaries
and found the remarkable result that general relativity may cause otherwise
stable stars to collapse prior to merging. In another recent paper
Cook et al. [3] tested the Wilson scheme for isolated, rapidly
rotating relativistic stars. They found a deviation from the Einstein theory
of at most 5\% which they interpreted as very encouraging for a better
understanding of binary star evolution.

In this paper we apply the Wilson scheme to point-like binary systems
at the second post-Newtonian approximation and calculate the periastron
advance as well as the orbital period. For circular motion, also the dependence
of the angular momentum on the orbital angular frequency is given.
In addition, we use the Hamiltonian
of the Wilson scheme to calculate a formal innermost stable circular orbit
(ISCO) for binary systems. The obtained results are confronted with the
corresponding results of the Einstein dynamics.

In the Einstein theory the periastron advance
and the orbital period have been calculated by
Damour and Sch\"afer [4] starting from a second post-Newtonian Hamilton
function. In isotropic coordinates and in the center of mass
system (${\bf{P}}_1=-{\bf{P}}_2={\bf{P}}$),
in reduced varibales (${\bf{p}}={\bf{P}}/\mu$, ${\bf{r}}={\bf{R}}/GM$)
the reduced Hamilton function $\hat H=H/\mu$ reads
\begin{eqnarray}
{\hat H}({\bf r}, {\bf p}, \nu)&=&\frac{1}{2}{\bf p}^2-\frac{1}{r}
-\frac{1}{8c^2}(1-3\nu){\bf p}^4-\frac{1}{2rc^2}[(3+\nu)
{\bf p}^2+\nu({\bf n}\cdot{\bf p})^2] \nonumber \\ &+&
\frac{1}{2r^2c^2}+\frac{1}{16c^4}(1-5\nu+5\nu^2){\bf p}^6 \nonumber
\\ &+&\frac{1}{8rc^4}[(5-20\nu-3\nu^2){\bf p}^4-2\nu^2{\bf p}^2
({\bf n}\cdot{\bf p})^2-3\nu^2({\bf n}\cdot{\bf p})^4]\nonumber
\\&+&\frac{1}{2r^2c^4}[(5+8\nu){\bf p}^2+3
\nu({\bf n}\cdot{\bf p})^2]-\frac{1}{4r^3c^4}(1+3\nu),
\label{2pNrh}
\end{eqnarray}
where $\nu=\mu/M$ with $\mu = M_1M_2/M$ and $M=M_1+M_2$. $M_1$ and $M_2$
denote the masses of the two bodies. The linear momenta of the bodies are
${\bf{P}}_1$ and ${\bf{P}}_2$, and $\bf{R}$ denotes a difference of their
coordinate position vectors, ${\bf{R}} = {\bf{R}}_1 - {\bf{R}}_2$;
$c$ is the velocity of light.

The fractional periastron advance per orbital revolution, $k$, and the orbital
period, $P$, were found to be
\begin{eqnarray}
k &=&\frac{3}{h^2c^2}\left[1+\frac{1}{2}(5-2\nu)
\frac{E}{c^2}+\frac{5}{4}(7-2\nu)\frac{1}{h^2c^2}\right],\label{2pnpe}\\
P&=&\frac{2\pi GM}{\sqrt{-2E}^3}\left[1-\frac{1}{4}(15-\nu)\frac{E}{c^2}-\frac{3}{32}(35+30\nu +3\nu ^2)
\frac{E^2}{c^4}\right. \nonumber \\
&+&\left. \frac{3}{2}(5-2\nu )\frac{\sqrt{-2E}^3}{hc^4}\right],
\end{eqnarray}
where $E$ is the total center-of-mass energy (numerical value of $\hat H$) and where
$h$ is the absolute value of the reduced angular momentum ${\bf{J}}/GM\mu$.

The Hamilton function in the Wilson scheme is easily obtained
as the two--body special case of the n-body matter Hamilton function $H_{mat}$
of Sch\"afer [5] (eq. (3.14)). In reduced form this Hamiltonian reads
(isotropic coordinates),
\begin{eqnarray}
{\hat H}_{mat}({\bf r}, {\bf p}, \nu)&=&\frac{1}{2}{\bf p}^2-\frac{1}{r}
-\frac{1}{8c^2}(1-3\nu){\bf p}^4-\frac{1}{2rc^2}[(3+\nu)
{\bf p}^2+\nu({\bf n}\cdot{\bf p})^2] \nonumber \\ &+&
\frac{1}{2r^2c^2}+\frac{1}{16c^4}(1-5\nu+5\nu^2){\bf p}^6 \nonumber
\\ &+&\frac{5}{8rc^4}(1-4\nu){\bf p}^4
+\frac{1}{4r^2c^4}[(10+19\nu){\bf p}^2-3
\nu({\bf n}\cdot{\bf p})^2]-\frac{1}{4r^3c^4}(1+\nu).
\label{2pNhmat}
\end{eqnarray}
In the test--body limit, $\nu = 0 $, the two Hamilton functions $\hat H$ and
$\hat H_{mat}$ are identical as they should be on reasons of spherical symmetry
of the central body.

The periastron advance of the dynamics eq.\,(\ref{2pNhmat}) is easily obtained following 
the path way of Ref.~[4]. It comes out in the form
\begin{eqnarray}
k_{mat} &=& \frac{3}{h^2c^2}\left[1+\frac{1}{2}(5-\frac{3}{2}\nu-\frac{41}{12}\nu ^2)
\frac{E}{c^2}+\frac{1}{4}(35-\frac{27}{2}\nu-\frac{31}{4}\nu ^2)\frac{1}{h^2c^2}\right].\label{pea1}
\end{eqnarray}

As one can see from equations (\ref{2pnpe}) and (\ref{pea1}) the two different 
periastron advances have the following structure
\begin{equation}
k=\frac{1}{c^2}k^{1pN}_o+\frac{1}{c^4}(k^{2pN}_o+k^{2pN}_{\nu}),
\end{equation}
where $k^{npN}_o$ denotes the $\nu$--independent terms and $k^{npN}_\nu$ 
the $\nu$--dependent terms of $k$. $k$ and $k_{mat}$ are different in the 
$\nu$--dependent terms of the second post--Newtonian order only.
Using the Newtonian relation between energy and angular momentum,
$E=(e^2-1)/2h^2$, where $e$ denotes the eccentricity of the binary orbit,
for the same energy and angular monentum, the fractional difference between 
the two periastron advances at the
genuine two--body 2pN level reads
\begin{equation}
\Delta k=(k - k_{mat}) /|k^{2pN}_\nu| = \frac{1}{24(4+e^2)}\left[4(12+13\nu )-
(6-41\nu )e^2\right].\label{dk}
\end{equation}
The application of this expression to the case of equal--mass binaries ($\nu=1/4$) 
in circular motion orbit ($e=0$) gives a fractional difference between 
$k$ and $k_{mat}$ at the genuine two--body 2pN level of about 63\%.

The orbital period in the Wilson scheme is obtained in the form, again
following the route of Ref.~[4],
\begin{eqnarray}
P_{mat}&=&\frac{2\pi
GM}{\sqrt{-2E}^3}\left[1-\frac{1}{4}(15-\nu)\frac{E}{c^2}-\frac{15}{32}(7+6\nu-25\nu^2)
\frac{E^2}{c^4}\right. \nonumber \\
&+&\left. \frac{1}{2}(15-\frac{9}{2}\nu-\frac{41}{4}\nu^2)\frac{\sqrt{-2E}^3}{hc^4}\right].
\end{eqnarray}
Analogously to the periastron advance, the two orbital periods
have the following structure
\begin{equation}
P=P^N+\frac{1}{c^2}(P^{1pN}_o+P^{1pN}_\nu)+\frac{1}{c^4}(P^{2pN}_o+P^{2pN}_\nu).
\end{equation}
Also they are different in the $\nu$--dependent terms of the second post--Newtonian
order only.
For the same energy and angular momentum, the fractional difference between the two
orbital periods at the genuine two--body 2pN level reads
\begin{equation}
\Delta P= (P - P_{mat}) /|P^{2pN}_\nu| =-\frac{16}{3}\frac{6-\nu\left
(41-24\sqrt{1-e^2}\right)}
{128+3(10+\nu)\sqrt{1-e^2}},\label{dP}
\end{equation}
where we have used again the Newtonian relation between energy and angular
momentum. Compared to $\Delta k$ in the case of $e=0$
and $\nu=1/4$, the expression (\ref{dP}) reaches 6\% only.

For circular orbits the angular frequency, $\omega$, is defined
through the expression $\Phi/P$, where $\Phi$, the angle advance fore one orbital
period, is given by $\Phi=2\pi(1+k)$. Taking into account the relation between
the energy and the angular momentum for circular orbits, for our two dynamical
situations, Einstein and Wilson respectively, the relation between angular momentum
and orbital angular frequency turn out to be
\begin{equation}
h=\frac{1}{\omega^{1/3}}\left[1+\frac{1}{2}(3+\frac{1}{3}\nu)\frac{\omega^{2/3}}{c^2}
+\frac{1}{8}(27-19\nu+\frac{1}{3}\nu^2)\frac{\omega^{4/3}}{c^4}\right]\,
\end{equation}
\begin{equation}
h_{mat}=\frac{1}{\omega^{1/3}}\left[1+\frac{1}{2}(3+\frac{1}{3}\nu)
\frac{\omega^{2/3}}{c^2}
+\frac{1}{8}(27-39\nu-\frac{17}{3}\nu^2)\frac{\omega^{4/3}}{c^4}\right].
\end{equation}
The fractional difference between $h$ and $h_{mat}$ at the genuine 
two--body 2pN level reads, applying the same frequency in both cases,
\begin{equation}
\Delta h= (h - h_{mat}) /|h^{2pN}_\nu| =6\frac{10+3\nu}{57-\nu}\,.\label{dh}
\end{equation}
For the case of equal masses, $\nu=1/4$, $\Delta h$ results in the value of 113.7\%.

\vspace{10mm}
\noindent
In modifying a procedure of Kidder et al. [6] we calculate now a formal
ISCO for the second post--Newtonian binary dynamics
of the Wilson scheme.
The idea of the method by Kidder et al. was to add to the equations of
motion of a test--mass in Schwarzschild spacetime all $\nu$--dependent terms of
the second post--Newtonian binary equations of motion.

We apply this method to the Hamiltonian (\ref{2pNhmat}) which
describes the Wilson approximate dynamics at the second post--Newtonian
approximation.

The reduced energy $\hat H_o$ of a test body in
Schwarzschild spacetime, in isotropic coordinates, reads
\begin{equation}
\frac{\hat H_o}{c^2}(r,{\bf
p})=\frac{1-1/2c^2r}{1+1/2c^2r}\sqrt{1+\left(1+\frac{1}{2c^2r}\right)^{-4}\frac{{\bf
p}^2}{c^2}}-1.
\end{equation}
Augmentation of this expression by the second post--Newtonian $\nu$--dependent terms, 
$\hat{H}_\nu$, of the Hmiltonian (\ref{2pNhmat}) yields the following so--called 
{\em hybrid approximation}
\begin{equation}
\hat{H}_*=\hat{H}_o+\hat{H}_\nu.
\end{equation}
For circular orbits this expression can be expressed as a function of $r$ and
the reduced angular
momentum $h$ (${\bf n}\cdot {\bf p}=0$, ${\bf p}^2=h^2/r^2$).
The radius of the ISCO is than obtained by the aid of the equations
\begin{equation}
\frac{\partial \hat{H}_*(r,h)}{\partial r}\quad = \quad 0 \quad = \quad \frac{\partial^2
\hat{H}_*(r, h)}{\partial r^2}.
\end{equation}
We have solved these equations numerically. For two equal masses, $\nu=1/4$,
we obtained for the ISCO
\begin{equation}
r_{mat}=6.82\,GM/c^2.\label{rmat}
\end{equation}
Starting from the second post--Newtonian two--body Hamiltonian (\ref{2pNrh}) the
ISCO has been calculated by Sch\"afer and Wex [7]. They used the same method and
they obtained $r=7.34 GM/c^2$. The result (\ref{rmat}) obtained from the
Hamiltonian (\ref{2pNhmat}) differs by $7.1\%$ from this value for $r$.

For an easy comparison of the two different dynamical situations the numerical values
of the energy, $\hat{H}$, and the angular momentum, $h$, both coordinate independent quantities,
are given in Table 1.\\[12pt]
{\bf Tab.\,1\,:}\,ISCO radii $r$ (unit $GM/c^2$) in isotropic coordinates and related
angular momentum $J$ (unit $\nu GM^2/c$) and energy $E$ (unit $\nu Mc^2$) obtained
(i), for a test body in Schwarzschild spacetime, TBSD, (ii), for the two--body 
hybrid 2pN Wilson approximate dynamics, HWAD, and (iii), for the two--body 
hybrid 2pN Einstein dynamics, HED,\\[12pt]
\begin{tabular}{|l|c|c|c|}
\hline
   & TBSD & HWAD & HED\\
\hline
\hline
$r$ & 4,95 & 6,82 & 7,34 \\
\hline
$J$    & 3,46 & 3,59 & 3,71 \\
\hline
$E$    & -0,057 & -0,050 & -0,047 \\
\hline
\end{tabular}

\vspace{10mm}

The results
obtained for the periastron advances, eqs. (2) and (5), and the orbital periods,
eqs. (3) and (8), are valid in dynamical regimes where post--Newtonian
appproximations apply. In the test--body limit as well as in the first
post--Newtonian approximation the periastron advances and the orbital periods
coincide. They only differ in the genuine binary 2pN
parts, this means in the $\nu$--dependent terms of 2pN order.
The fractional difference of the 2pN $\nu$-dependent terms for $h$ takes the
remarkably large value of about 114\% in the case of equal--mass binaries 
in circular motion with the same orbital frequency.

On the other side, the estimated value for the ISCO
of the hybrid 2pN Wilson scheme, eq. (14), differs by about 7\% from
the previously calculated value for the 2pN binary dynamics, (\ref{2pNrh}), by
Sch\"afer and Wex [7].
In Tab.~1 we have summarized the numerical values for the ISCO radius, the energy
and the angular momentum for the different models. Surely, near the last stable
circular
orbit post--Newtonian approximations loose their meaning. Nevertheless, it
seems reasonable to conclude that in the Wilson scheme the binary system, near the
ISCO, is stronger bounded (see Tab.~1). The conclusion is supported by the fact the
in regimes where the 2pN approximation applies the stronger binding holds, see 
Fig.~1\,: Given the same orbital angular frequency, the Wilson approximate 
dynamics has smaller angular momentum, i.e. smaller moment of inertia. Another 
support results from the energy expression as function of $\omega$, most easily
obtained from the relation $dE=\omega dh$, also showing stronger binding.

\vspace{8mm}
{\bf References:}\\[12pt]
\noindent
[1] J.R. Wilson, in: {\it Texas Symposium on 3-Dimensional Numerical
Relativity}, edited by Matzner (University of Texas, Austin, Texas, 1990);
J.R. Wilson and G.J. Mathews, in: {\it Frontiers in Numerical Relativity},
edited by C.R. Evans, L.S. Finn., and D.W. Hobill (Cambridge University Press,
Cambridge, England, 1989), pp. 306-314.\\[12pt]
[2] J.R. Wilson and G.J. Mathews, Phys. Rev. Lett. {\bf 75}, 4161 (1995);
J.R. Wilson, G.J. Mathews and P. Marronetti, {\it Relativistic numerical
model for close neutron star binaries}, gr-qc/9601017, 4 Mar 1996, submitted
to Phys Rev. D (1996).\\[12pt]
[3] G.B. Cook, S.L. Shapiro, and S.A. Teukolsky, {\it Testing a simplified
version of Einstein's equations for numerical relativity}, gr-qc/9512009,
5 Dec 1995.\\[12pt]
[4] T. Damour and G. Sch\"afer, Nuovo Cimento B {\bf 101}, 127 (1988).\\[12pt]
[5] G. Sch\"afer, Ann. Phys. (N.Y.) {\bf 161}, 81 (1985).\\[12pt]
[7] L.E. Kidder, C. M. Will and A. G. Wiseman, Class. Quantum Grav. {\bf 9}, L125 (1992);
also see Phys. Rev. D {\bf 47}, 3281 (1993).\\[12pt]
[8] N. Wex and G. Sch\"afer, Class. Quantum Grav. {\bf 10}, 2729 (1993);
G. Sch\"afer and N. Wex, in:
{\it Perspectives in Neutrinos, Atomic Physics and Gravitation},
{\it Proceedings of the XIIIth Moriond Workshop}, edited by J. Tr$\hat{\rm a}$n 
Thanh V$\hat{\rm a}$n,
T. Damour, E. Hinds, and J. Wilkerson (1993), p. 513--517.\\[12pt]

\vspace{2cm}
\psfig{figure=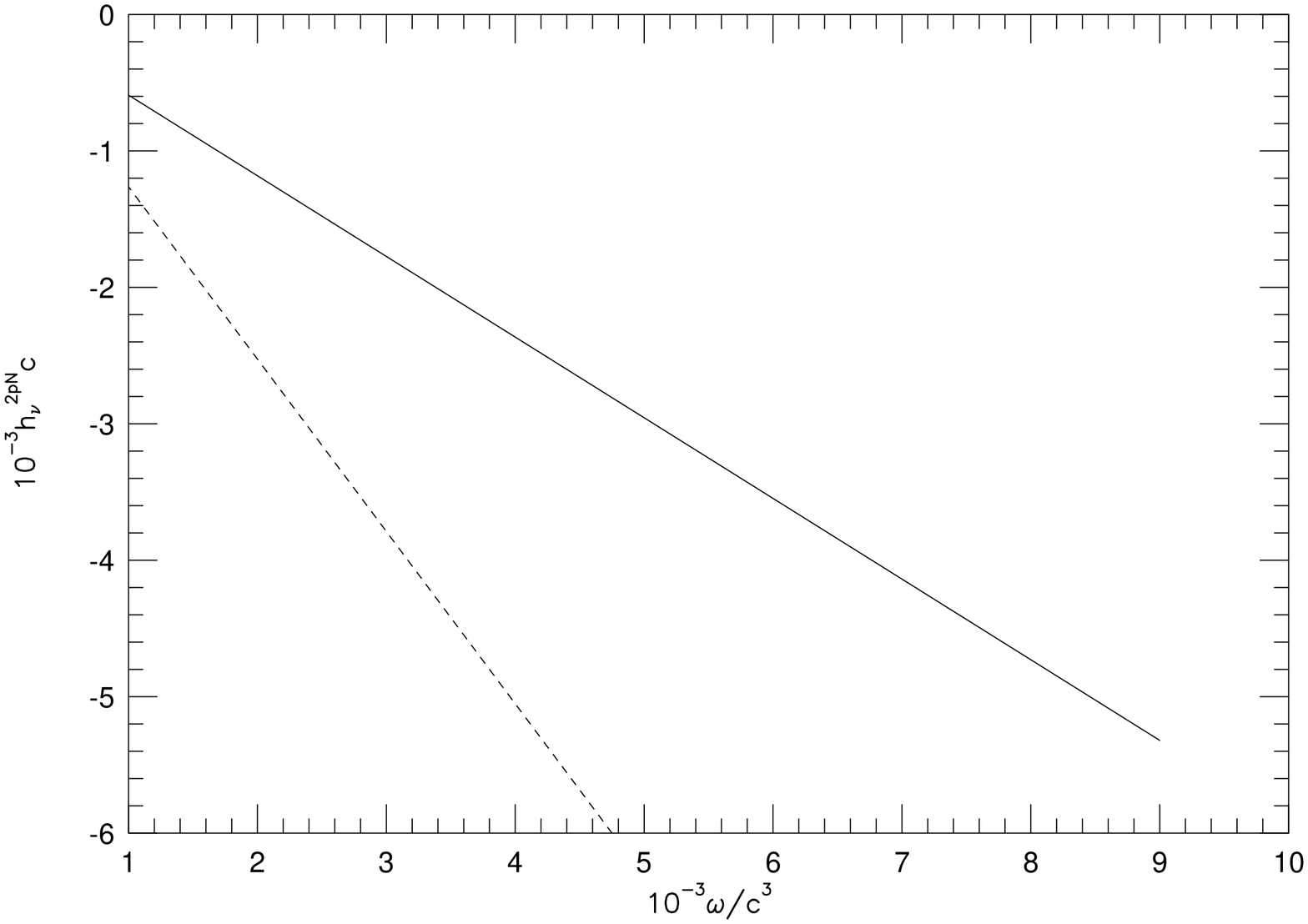,height=110mm,width=180mm}
\vspace{5mm}

\noindent
{\bf Fig.\,1\,:\,}The value of the genuine two--body 2pN term in $h$
is plotted as function of $\omega$, the orbital angular frequency
for circular motion, for the
both models. The upper curve belongs to the 2pN
Einstein dynamics whereas the lower curve belongs to the 2pN Wilson approximate
dynamics.
\end{document}